%% file: gaspi_ft.tex
\documentclass[letterpaper]{IEEEtran}
\usepackage{mathptmx}
\usepackage{graphicx}
\usepackage{hyperref}
\usepackage{breakurl}
\usepackage[english]{babel}
\usepackage{color}
\usepackage{listings}
\usepackage{subfig}
\usepackage{array}
\usepackage{cite}
\usepackage{algorithm}
\usepackage{algpseudocode}
\usepackage{hhline}
\definecolor{mygray}{rgb}{0.5,0.5,0.5}
\newcolumntype{L}[1]{>{\raggedright\let\newline\\\arraybackslash\hspace{0pt}}m{#1}}
\newcolumntype{C}[1]{>{\centering\let\newline\\\arraybackslash\hspace{0pt}}m{#1}}
\newcolumntype{R}[1]{>{\raggedleft\let\newline\\\arraybackslash\hspace{0pt}}m{#1}}

\title{Building a fault tolerant application using \\ the GASPI communication layer}

\author{\IEEEauthorblockN{Faisal Shahzad\IEEEauthorrefmark{1}, Moritz Kreutzer\IEEEauthorrefmark{1}, Thomas Zeiser\IEEEauthorrefmark{1}, Rui Machado\IEEEauthorrefmark{2},\\ 
Andreas Pieper\IEEEauthorrefmark{3}, Georg Hager\IEEEauthorrefmark{1} and Gerhard Wellein\IEEEauthorrefmark{1}}\\
\IEEEauthorblockA{\IEEEauthorrefmark{1}Erlangen Regional Computing Center, University of Erlangen-Nuremberg Erlangen, Germany}\\
\IEEEauthorblockA{\IEEEauthorrefmark{2}Fraunhofer Institute for Industrial Mathematics ITWM, Fraunhofer Platz 1, Kaiserslautern, Germany} \\
\IEEEauthorblockA{\IEEEauthorrefmark{3}Institute of Physics, University of Greifswald, Greifswald, Germany}
}

\begin{document}
\maketitle
\begin{abstract}

It is commonly agreed that highly parallel software on Exascale computers 
will suffer from many more runtime failures due to the decreasing 
trend in the mean time to failures\,(MTTF). Therefore, it is not surprising 
that a lot of research is going on in the area of fault tolerance and fault 
mitigation. Applications should survive a failure and/or be able to recover 
with minimal cost. MPI is not yet very mature in handling failures, the 
User-Level Failure Mitigation\,(ULFM) proposal being currently the most 
promising approach is still in its prototype phase.

In our work we use GASPI, which is a relatively new communication
library based on the PGAS model. It provides the missing features to
allow the design of fault-tolerant applications. Instead of
introducing algorithm-based fault tolerance in its true sense, we
demonstrate how we can build on (existing) clever checkpointing and
extend applications to allow integrate a low cost 
fault detection mechanism and, if necessary,
recover the application on the fly. The aspects of process
management, the restoration of groups and the recovery mechanism is presented
in detail. We use a sparse matrix vector multiplication based
application to perform the analysis of the overhead introduced by such 
modifications. Our fault detection mechanism causes no
overhead in failure-free cases, whereas in case of failure(s), the
failure detection and recovery cost is of reasonably acceptable 
order and shows good scalability.

\end{abstract}

{\bf Keywords:} GASPI; GPI; fault detection; fault recovery/tolerance; checkpoint-restart; pre-allocated spare processes 

\input{introduction}
\input{background}

%
%
\input{gaspi_intro}

%
\input{proposed_technique}

%
\input{experimental_framework}

%
\input{results}

%
\input{related_work}

%
\input{summary}


\section*{Acknowledgment} 
This work was partly supported by the German Research
Foundation\,(DFG) through the Priority Programme 1648 ``Software for
Exascale Computing"\,(SPPEXA) and partly by Federal Ministry of
Education and Research\,(BMBF) under project ``A Fault Tolerant
Environment for Peta-scale MPI-solvers"\,(FETOL) (grant
No. 01IH11011C).


\bibliographystyle{IEEEtran}
\bibliography{gaspi_ft.bib}

\end{document}

%% file: introduction.tex
\section{Introduction}
The advances in computational capacity of HPC clusters have enabled
many fields in research and industry to progress far beyond
imagination. Still the demand of more computational capacity is never
ending. In the recent past, the consistent exponential growth is
achieved with the help of extreme levels of hardware parallelism. This
causes a severe reduction in mean time to failure (MTTF) of the
overall systems and is visible with every new generation of large
clusters. For example, the `Intrepid' BlueGene/P system (debuted as
$\#$ 4 on the top500 list of June 2008 \cite{website:top500}, installed at
the Aragonne National Laboratory) is reported to have the MTTF of 7.5
days \cite{addressing-failures-inexacale}. In contrast, a more recent
BlueGene/Q cluster `Sequoia' (debuted as \# 3 according to Nov. 2013
list, installed at Lawrence Livermore National Laboratory) has a node
failure rate of 1.25 per day \cite{Invited_talk_Jackdongara}.  On the
way to exascale machines, the MTTF is expected to reduce to the order
of hours or minutes \cite{Inter-AgencyWorkshop,
  schroeder-understanding-failures}. This indicates an alarming
behavior which, if not taken care of, will question the usability of
clusters at exascale.

In HPC systems, programs can face many kinds of failures during
runtime, e.g., hardware and software faults, silent errors, Byzantine
failures, etc.  According to \cite{El-Sayed:2013:RLF_Bianca}, 60\% of
all failures are attributed to either memory or CPU failures.  Such
failures, in addition to others, eventually lead to process or
node-level failures, which are the focus of this work. The Majority of the
literature regarding fault tolerance towards fail-stop failures falls into
either one or a combination of the following four categories:
algorithm based fault tolerance\,(ABFT), checkpoint/restart\,(C/R),
message logging, and redundancy \cite{hursey-phd-thesis-2010}.

So far most algorithms (and underlying communication models) are built
under a comprehensible assumption that the communication partners of
every process stay alive and functional during the entire run of the
program. Consequently, the failure of even a single process leads to
the failure of the whole application. For large scale applications, it is
beneficial to drop this assumption. This would mean that the program should
continue to run even after the failure of a certain amount of processes, i.e., 
node failures. This opens a new dimension of research in the
field of fault tolerance. In this context, the first step is to build/utilize
a communication library that can provide the necessary supporting
functionalities for this purpose, i.e., health information of processes,
failure detection mechanisms, propagation of failure information to all
relevant processes, etc. During the development of MPI 3.0, efforts
were underway to introduce process-level failure tolerance
\cite{bland2012proposal}, but were not eventually successful. The 
fault tolerant working group in the MPI forum is currently working on the
User-Level Failure Mitigation\,(ULFM) standard proposal and its
prototype implementation (based on OpenMPI) for its potential
inclusion in the MPI 3.1 standard \cite{Bland_post_failure}.  Despite the
apparent attractiveness of this new approach, it comes with a new set
of challenges, i.e., correct and consistent failure acknowledgment,
rebuilding communication infrastructure, recovering lost data from
failed process(es), etc. These building blocks require extra effort and
caution during application development stages.

In this work, we use the GPI-2 \cite{website:gpi2} implementation of
the GASPI specification \cite{website:gaspi} to build a fault tolerant
application that is capable of recovering dynamically from process
failures.

The main contributions of our work are as follows:
\begin{enumerate}
\item Design and implementation of the fault detection\,(FD) mechanism
  to detect fail-stop failures in parallel applications. The recovery
  mechanism is then implemented, which includes the repairing
  of broken communication infrastructure and the recovery of the lost
  data of the failed processes.
\item The FD mechanism is applied to a sparse matrix vector
  multiplication \,(spMVM) based application, namely Lanczos eigensolver, 
  to showcase its real usage. 
  The necessary algorithmic restructuring is presented in
  detail. The changes in the underlying spMVM library are also
  highlighted.
\item For recovering the data of the failed processes, the
  application-level C/R approach is used. Despite
  the criticism it faces, the recent neighbor node-level C/R
  optimizations \cite{Moody:2010:multi-level-checkpointing-system,
    Sato:2012:DMN:SCR} have enabled it to be a good candidate
  on future exascale systems \cite{Inter-AgencyWorkshop}. We have
  implemented a neighbor node-level C/R library for GPI-2
  applications. In case of a node failure, the neighboring
  nodes are bound to change, thus the C/R library is
  also made fault aware.
\item A benchmark study has been performed to analyse the overheads
  involved in this fault tolerance approach. This includes a range of
  application runtime cases, i.e., overhead in failure-free case and 
  overhead with 1,2,3 failure recoveries, etc. Moreover, the scalibility of the 
  fault detection method is tested with up to 256 nodes.
\end{enumerate}

With this practice, we have developed a fault tolerant application
which can heal itself dynamically after the loss of one or more
processes. Thus the time for restarting the job manually and wait-time
spent in the queueing system for a new job request are 
avoided. The concept can be applied to other applications with
different communication libraries as well when they support fault
tolerance. We think that this is a good starting point to gain
experience and become aware of the challenges involved and their
potential solutions for utilization of such fault tolerant
communication libraries.

The paper follows the following structure. Some preliminary
definitions and concepts are described in
Sect. \ref{sec:background}. In Sect. \ref{sec:gaspi_intro}, the GASPI
communication interface is briefly introduced along with its fault
tolerance features.  The design and implementation of our fault tolerance
method along with the fault detection mechanism are described in
Sect. \ref{sec:design_and_implementation}. Section
\ref{sec:experimental_framework} introduces the benchmark environment
and provides application specific details for benchmarks. The overhead
analysis of our implementation is presented with benchmarks in
Sect. \ref{sec:results}. A brief summary of the related work is
described in Sect. \ref{sec:related_work}. We conclude the paper in
Sect. \ref{sec:summary} with a short discussion about the challenges of such 
a fault tolerance approach and possible improvements.

%% file: background.tex
\section{Preliminaries}
\label{sec:background}

There are three central components in application-driven
fault tolerance. In this paper, we use the term `application-driven'
fault tolerance for an approach where an application can heal itself
dynamically despite one or more process failures. 
The first and foremost component is to have a consistent, accurate and
reliable failure detection mechanism. Design and implementation
of a fault detector for asynchronous systems is a complex task. A
distributed system is termed as `asynchronous' if there is no upper
bound on message transmission delays and process execution time, and
all real HPC systems fall into this category.  There are two basic
properties of a fault detector \cite{Chandra:1996:UFD:226643.226647}:
1) Completeness: The crashed processes are suspected by some healthy
processes, after a certain amount of time. A strong completeness
implies that every failed process is eventually detected by every
healthy process. 2) Accuracy: Every detected failure corresponds to a
crashed process (no false-positives). The complete satisfaction of both these
properties is theoretically impossible 
\cite{Chandra:1996:UFD:226643.226647}. Thus most FD
implementations are willing to tolerate some level of inaccuracy but
require strong completeness.

The implementation of ULFM relies on detection of faults based on the
communication failure between processes. After a failed process is
detected, the user can revoke the rest of the communication in a
communicator and propagate the failure information about failed processes 
using ULFM functions.
In this work, we choose a different approach where an explicit fault detector 
monitors the health of every other process. Due to the PGAS nature of
communication in GASPI, this brings certain advantages in terms of
developing consensus between processes after failure(s) and incurs no
overhead in the failure-free case ({discussed in detail in
Sect. \ref{sec:design_and_implementation}).

After a correct failure detection, the second step involves the
selection of a communication reconstruction and domain redistribution strategy. 
There can be two approaches in this context
\cite{evaluating-ULFM-for-mpi-applications}.  1) Shrinking recovery:
In this method, the application proceeds with the rest of available
processes after failure(s) and requires redistribution of domain.
2) Non-shrinking recovery: This method involves using new process(es) to replace
the failed one(s), where the work distribution of the application is not
changed after failures. Depending on the application and the
communication library, either of these techniques can be
beneficial. In our work, we have used a non-shrinking recovery
method.

The third stage involves the data recovery of the lost process(es)
and continuing further computation. Data recovery can either be done
by reading a checkpoint or by using an algorithm based fault
tolerance\,(ABFT) approach. An ABFT approach is by nature highly
algorithm-specific and cannot be generalized; therefore, we have
opted for application specific node-level C/R approach in our
application.

%% file: gaspi_intro.tex
\section{Introduction to GASPI and GPI-2}
\label{sec:gaspi_intro}
GASPI is a communication library for C/C++ and Fortran. It is based on
a PGAS style communication model where each process owns a partition
of a globally accessible memory space. GPI-2, the GASPI
implementation, takes full advantage of the hardware capability to
perform remote direct memory access\,(RDMA). More importantly, there
is a focus on providing truly asynchronous communication to overlap
computation and communication, and on thread-safe communication which
allows multi-threaded applications with a fine-grained communication
and asynchronous execution capability.

GASPI defines a very compact API consisting of one-sided communication
routines, passive communication, global atomics and collective
operations. It also defines groups which are similar to MPI
communicators and are used in collective operations. Furthermore,
there is the concept of segments. Segments are contiguous blocks of
memory and can be made accessible (to read and write) to all threads
on all ranks of a GASPI program. Data to be communicated is thus
placed in such segments.

Given the aforementioned increasing need for fault tolerant
applications, GASPI was designed with that in mind. It supports
application-driven fault tolerance on the process level. This
means that the failure of a single process does not cause the entire
program to fail. The program can in fact react to the failure and try
to recover by ``healing itself". This is achieved with two simple
concepts: timeouts and an error state vector.

GASPI provides a timeout mechanism to all potentially blocking
procedures. The user provides a timeout (in milliseconds) and the
procedure returns after that time if it could not successfully
complete. Furthermore, since a timeout does not necessarily imply an
error or failure from the remote part, GASPI provides an error state
vector that holds the state of processes. The state vector is set
after every erroneous, non-local operation and can be used to detect
failures on remote processes. Currently, each rank can either have
a state of GASPI\_STATE\_HEALTHY or GASPI\_STATE\_CORRUPT. This error
state vector can be queried by the application using
\textit{gaspi\_state\_vec\_get} to determine the state of a remote
partner in case of timeout or error.

Along with the previous mechanisms, for the context of our work, we
have extended GPI-2 to provide an extra mechanism (not included in
the GASPI specification) for fault-tolerant applications. The
procedure \textit{gaspi\_proc\_ping} tests the availability of a particular
GPI-2 rank. As the name indicates, a \emph{ping} message is sent to a
particular process. In case a problem is detected, a GASPI\_ERROR is
returned to which the application can react.

In the next sections, we describe our technique in more detail.

%% file: proposed_technique.tex
\section{Design and Implementation}
\label{sec:design_and_implementation}
This section explains the details of our fault detector\,(FD), 
its implementation in the algorithm, and the data recovery
technique for our implementation. The basic idea behind our
implementation is to designate some processes as `idle processes' at
the start of the computation to facilitate non-shrinking recovery. The
remaining processes form the `worker group' and do
computation. The idle processes serves two purposes. One of the
pre-determined idle processes serves as a failure detector
process. The rest of the idle processes stay idle until FD detects a
failure and asks idle process(es) to act as rescue process(es) and
join the worker group.

\subsection{Failure Detection and Acknowledgment}
\label{subsec:failure_detection}
\begin{figure}
\centering
	\subfloat[Ping based fault detection by FD process]{%
	\includegraphics[width=.45\linewidth]{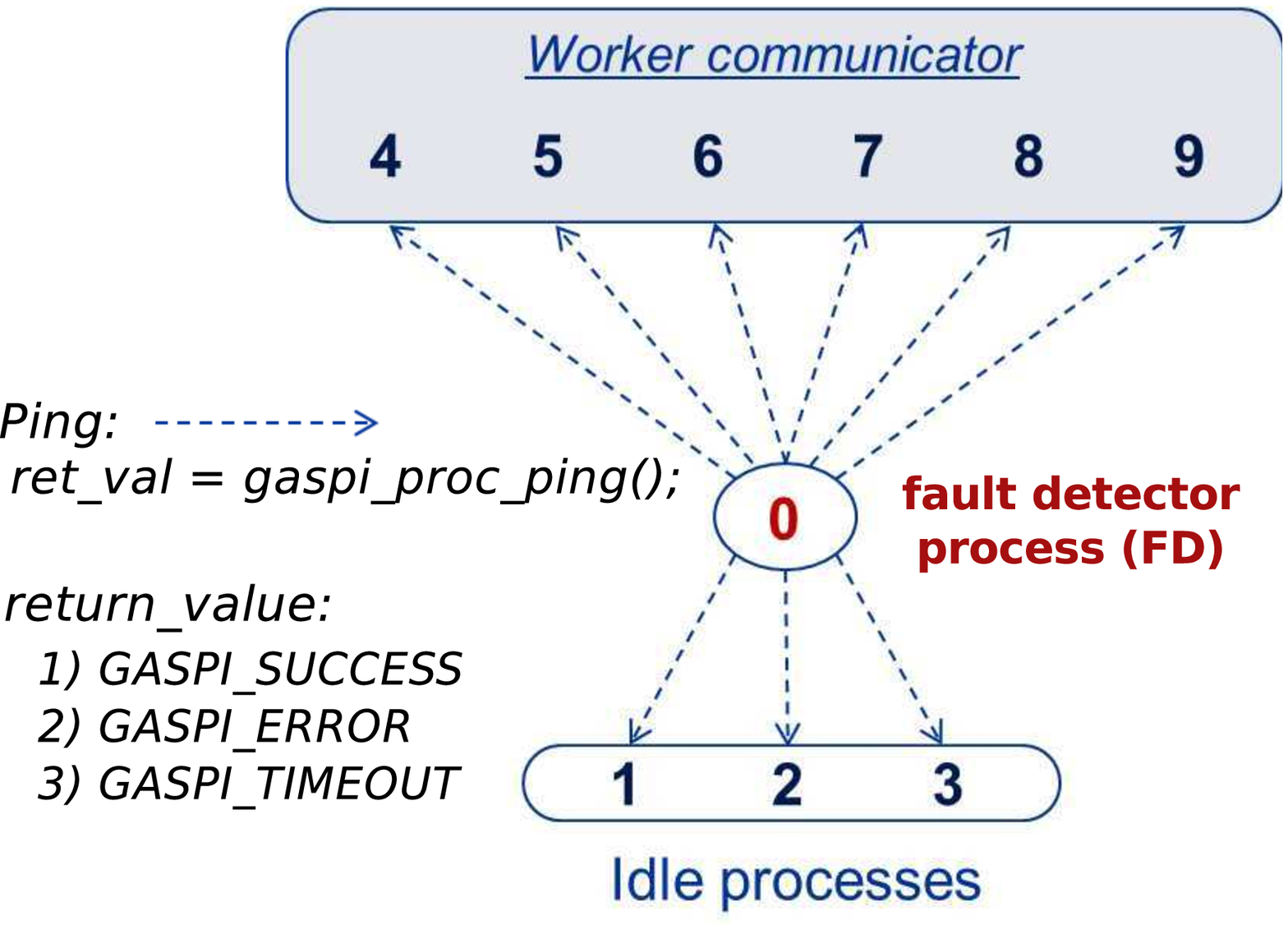}
	\label{fig:FD_ping_a}}
	\quad\quad
	\subfloat[FD detects failure of 6,7 processes and designates 1,2 as rescue processes]{%
	\includegraphics[width=.38\linewidth]{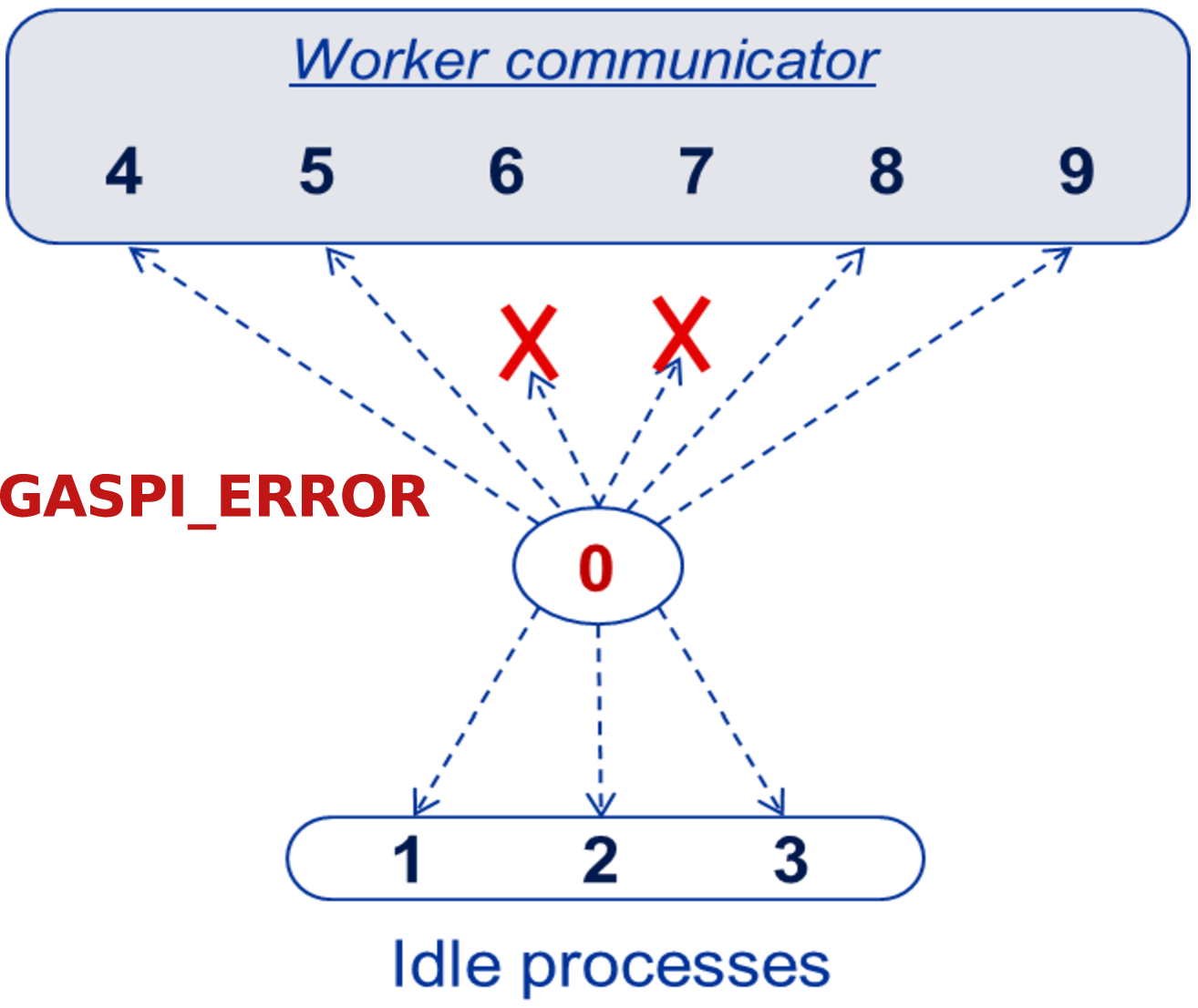}
	\label{fig:FD_ping_b}}
	\quad\quad
	\subfloat[The rescue processes (1,2) join the worker group during failure recovery stage]{%
	\includegraphics[width=.38\linewidth]{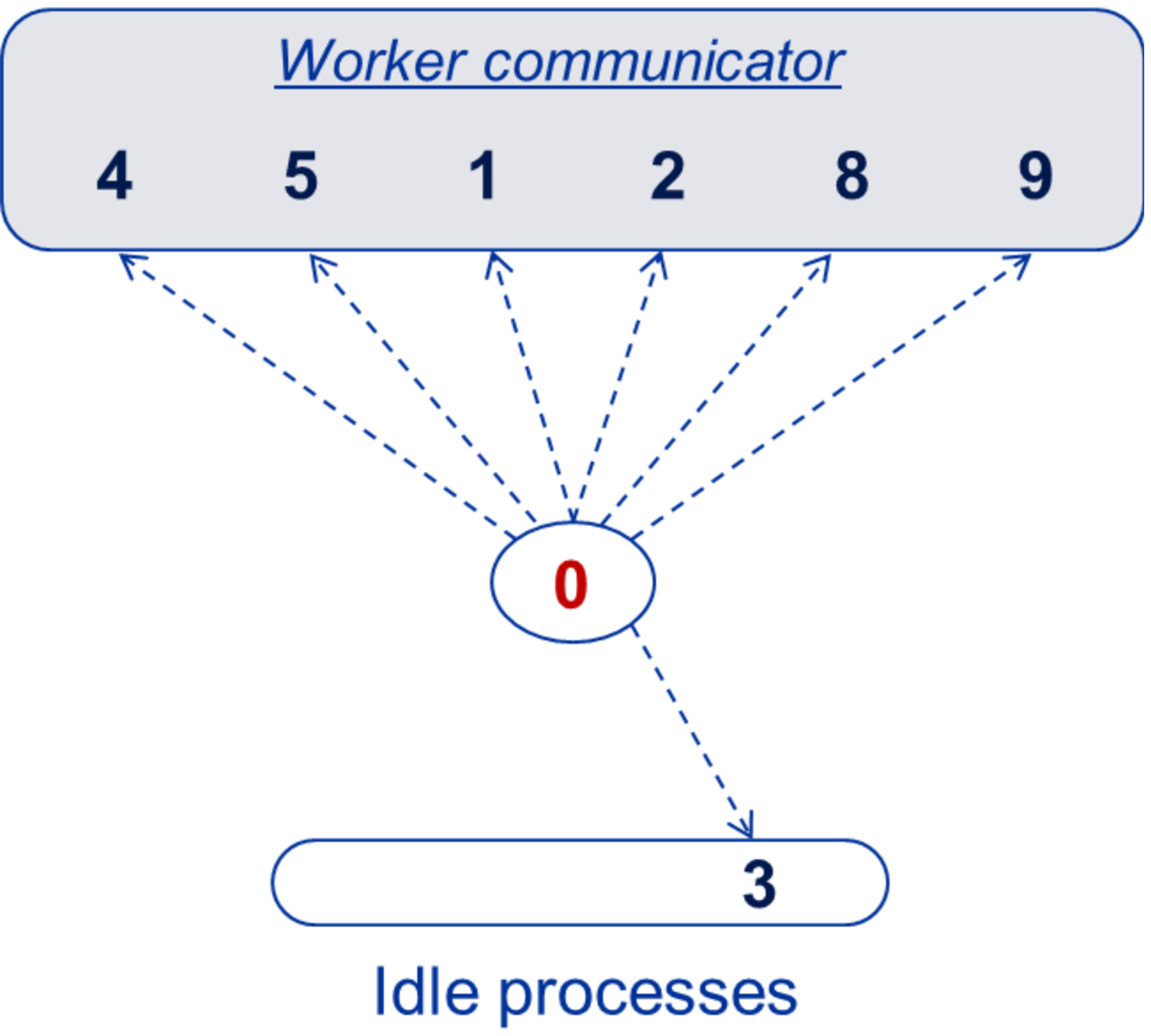}
	\label{fig:FD_ping_c}}
      \caption{The working of the fault detector process. The return
        value of ping is GASPI\_SUCCESS for all healthy processes
        (a), whereas it returns with GASPI\_ERROR
        for failed processes (b). After failure
        detection, FD informs all processes about failed processes as
        well as their replacement processes. A new worker group
        is then built with the help of rescue processes
        (c).}
	\label{fig:FDprocess}
\end{figure}

Our fault detection method relies on a dedicated FD process 
periodically checking
the health (i.e., aliveness) of all other processes. The FD process
pings every other process via \textit{gaspi\_proc\_ping()} and checks
for its return value as shown in Listing \ref{lst:glo_health_chk}. A
successful completion (i.e., return value of GASPI\_SUCCESS) of the
ping operation implies a healthy process. Figure \ref{fig:FD_ping_a}
shows the schematic diagram of such one-sided ping based
fault-detection. A rational ping frequency can be set by the user 
depending on the number of processes. 
After completion of a single cycle around all healthy
processes, the FD has an updated global health view. We rely on the
fact that each fail-stop failure eventually results in breakage of the
communication channel between the FD and the failed process. This
consequently leads to the return of ping operation with GASPI\_ERROR
(shown in Fig. \ref{fig:FD_ping_b}).  After detection of failed
process(es), the FD process informs all healthy processes about the
failed processes as well as their corresponding rescue processes. This
is done via one-sided write in the global memory of all healthy
processes.  Meanwhile, the worker processes communicating directly
with the failed processes keep on returning with GASPI\_TIMEOUT unless
a failure acknowledgment is received. The remaining healthy working
processes continue with their work until they also receive a failure
acknowledgment signal from the FD process. After failure acknowledgments,
no further regular application communication is performed and
processes enter the recovery stage of the algorithm. A threaded
implementation of the FD process enables it to monitor the health state of
multiple processes simultaneously. For exascale systems, 
multiple FDs can be used to distribute the health check scan of all 
processes.

\begin{lstlisting}[float=tb, caption=The global health check routine. The fault detector process periodically checks the health of all healthy processes., label=lst:glo_health_chk]
int glo_health_chk(gaspi_rank_t * failed_proc_list, gaspi_rank_t * rescue_proc_list, gaspi_rank_t *avoid_list){
	int comm_state = WORKING;
	gaspi_return_t retval;
	for(int rem_id=0; rem_id<numprocs(); ++rem_id){
		if(avoid_list[rem_id] != 1){	// protects messaging already discovered failed processes 
			retval = gaspi_proc_ping(rem_id, GASPI_BLOCK);
		}
		if(retval == GASPI_ERROR){
			avoid_list[rem_id]=1;
			comm_state = BROKEN;
		}
	}
	// informing all healthy processes about failed processes and their rescue processes.
	if(comm_state == BROKEN){
		make_failed_and_secue_proc_list(failed_proc_list, rescue_proc_list);
		report_healthy_processes_about_failure(failed_proc_list ,rescue_proc_list);
	}
	return comm_state;
}
\end{lstlisting}

\paragraph{Fault Detector Properties}
The FD process gets an updated global view of all processes' health at
the end of each ping cycle around all healthy processes. Thus, a
failed process is bound to be detected by the FD process in a finite
amount of time. The information about failed processes is then given to
all remaining healthy processes. Thus the FD process satisfies the
property of completeness. As a rare case, there can be instances that (due to a
network related problem) a healthy process is not reachable by the FD
process but is accessible by some or all other processes. This will
result in a false-positive signal about a processes failure. Thus the
property of accuracy is only loosely satisfied. A false-positive signal 
will lead the application to undergo the failure recovery phase but it does 
not affect the correctness of the program results.

\paragraph{Alternate investigated failure detection methods}
We also investigated the following failure detection mechanisms:
\begin{enumerate}
\item{Ping-based all-to-all: In this method each process performs
    periodic all-to-all pings to detect the failures.}
\item{Ping-based neighbor level: Each process `i' periodically pings
    only its neighboring process `i+1'. In case a failure is detected
    at the neighbor level, a ping based all-to-all operation is
    triggered on all processes to get a global health view.}
\end{enumerate}
In both of these approaches, a potential deadlock can arise in the cases
where multiple processes detect different sets of failed processes.
Reaching a consensus about the identified failures adds further
complexity in the algorithm. Moreover an all-to-all ping based
approach is not a scalable failure detection method and in failure
free cases a certain amount of overhead in both cases is introduced.
In contrast, a dedicated FD process with one-sided ping eliminates
the potential deadlock situation and causes negligible overhead in
failure-free cases.

\subsection{Communication reconstruction}
The FD process propagates the list of failed processes and the list of
respective rescue processes to all worker and rescue processes.  The
worker group is recreated in such a way that the failed processes are
left out and the rescue processes are included. Figure
\ref{fig:FD_ping_b} and \ref{fig:FD_ping_c} show the schematic of
failure acknowledgment and reconstruction of a working group after
failure. The steps involved in this recovery mechanism to
form a working group are shown in Listing
\ref{lst:reconstruct_comm_worker}.  The cases of transient failures
and a false-positive detected failures are handled using
\textit{gaspi\_proc\_kill} in the communication reconstruction phase.
It explicitly enforces the processes to die even if they were alive,
preventing them from further participation in the application.
\begin{lstlisting}[float=tb, caption=The reconstruction of the working group is performed in case a failure is detected in the health check routine., label=lst:reconstruct_comm_worker]
// check for failure acknowledgement signal
if( check_failed_proc_list(failed_proc_list) == true ){
	bool am_i_rescue_proc = am_i_rescue_process (rescue_proc_list);
	if(am_i_rescue_proc == true){
		update_myrank_active(myrank_active, gm_myrank_active);
		// rescue processes overtake the identity of the failed processes
	}
	if(am_i_rescue_proc == false){
		gaspi_group_delete(COMM_MAIN);
	}
	for(int i=0; i<num_failed_procs ; ++i){		
		gaspi_proc_kill(failed_proc_list[i], GASPI_BLOCK);
	}

	gaspi_group_t COMM_MAIN_NEW;
	gaspi_group_create(&COMM_MAIN_NEW);
	gaspi_number_t gsize;
	refresh_status_proc(status_processes);
	// status_processes stores information about state of processes (working, failed or idle). Also managed by FD process
	for(gaspi_number_t i = 0; i < numprocs(); i++){
		if(status_processes[i]==WORKING){	
			gaspi_group_add(COMM_MAIN_NEW, i);
			gaspi_group_size(COMM_MAIN_NEW, &gsize);
			if(gsize==numprocs_working)
				break;
		}
	}
	gaspi_group_commit (COMM_MAIN_NEW, GASPI_BLOCK);
	COMM_MAIN = COMM_MAIN_NEW;
	re_init_data_stucture_read_CP();
}
\end{lstlisting}

\subsection{Data Recovery: Neighbor node-level checkpoint/restart}
\begin{figure}[tb]
	\centering
	\includegraphics[width=8cm]{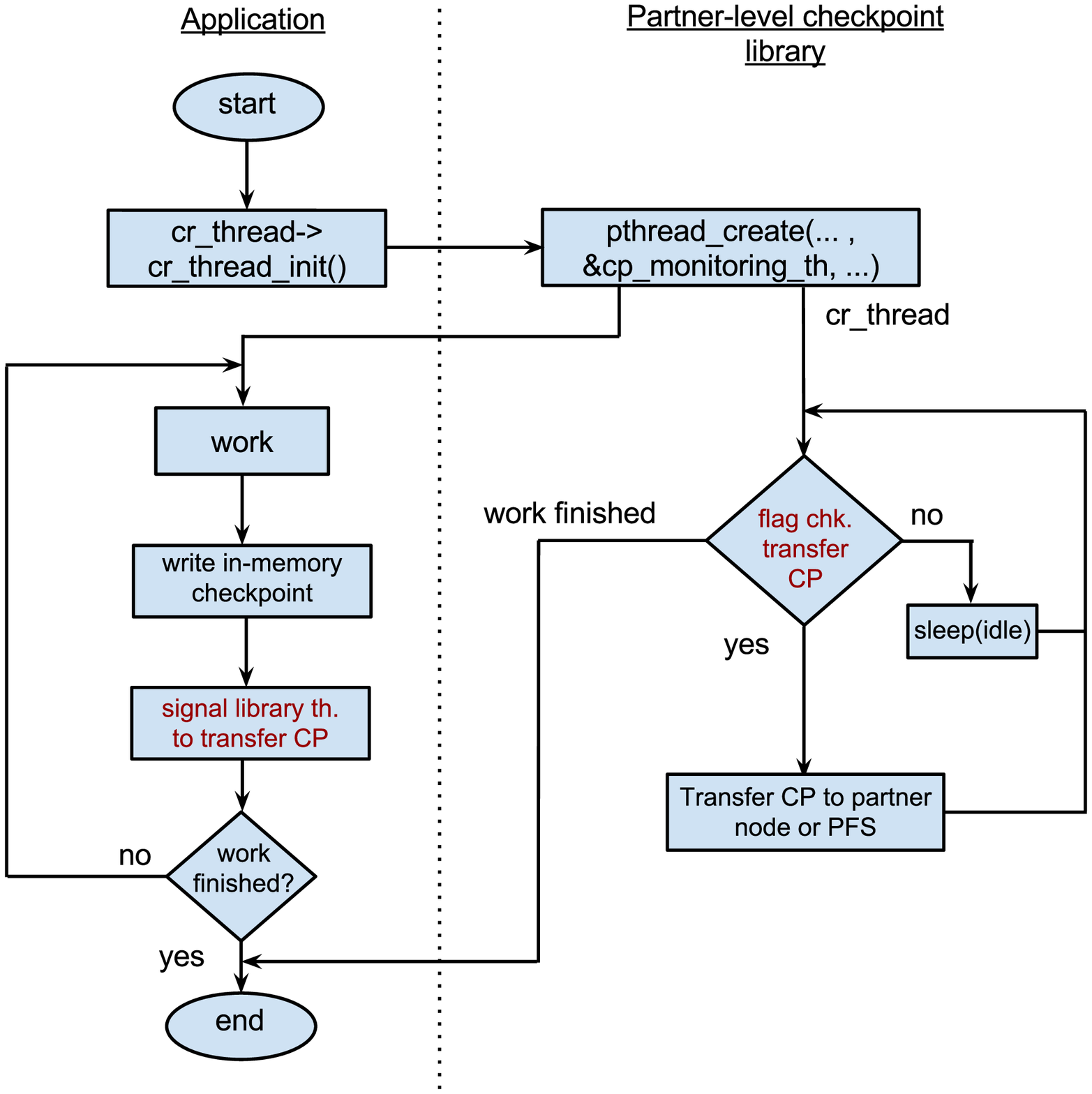}
	\caption[]{Schematic of the interaction between application and the neighbor node-level checkpointing library.
	\label{chart:flowchart-FT-library}}\vspace{.00cm} 
\end{figure}

We have utilized the classic application specific C/R technique for
the purpose of data recovery in our application. Typically checkpoints
are written to the parallel file system\,(PFS). Writing and retrieving
them from PFS is expensive. Therefore we have implemented a library
for GPI-2 to take checkpoints at neighbor node-level. Figure
\ref{chart:flowchart-FT-library} shows the schematic of an application
utilizing the library. At the \textit{init} call from the application,
the library creates a thread which waits for a signal from the
application. At a checkpoint iteration, the application first creates
a checkpoint on its local node and signals the library thread after
completion. The library thread then copies the local copy of the
checkpoint to the neighboring nodes. For a higher degree of
reliability, the user can also opt to make infrequent PFS-level copies
of the checkpoints. Due to the fault-tolerant nature of the algorithm,
the neighboring node of processes can change after failure
recovery. This requires the neighbor node-level C/R library to be
fault aware as well. After the reconstruction of the worker group,
the C/R library refreshes its list of neighboring processes based on
the failed processes list provided by the application thread.
\subsection{Application flow}
\begin{figure}[tb]
	\centering
	\includegraphics[width=8cm]{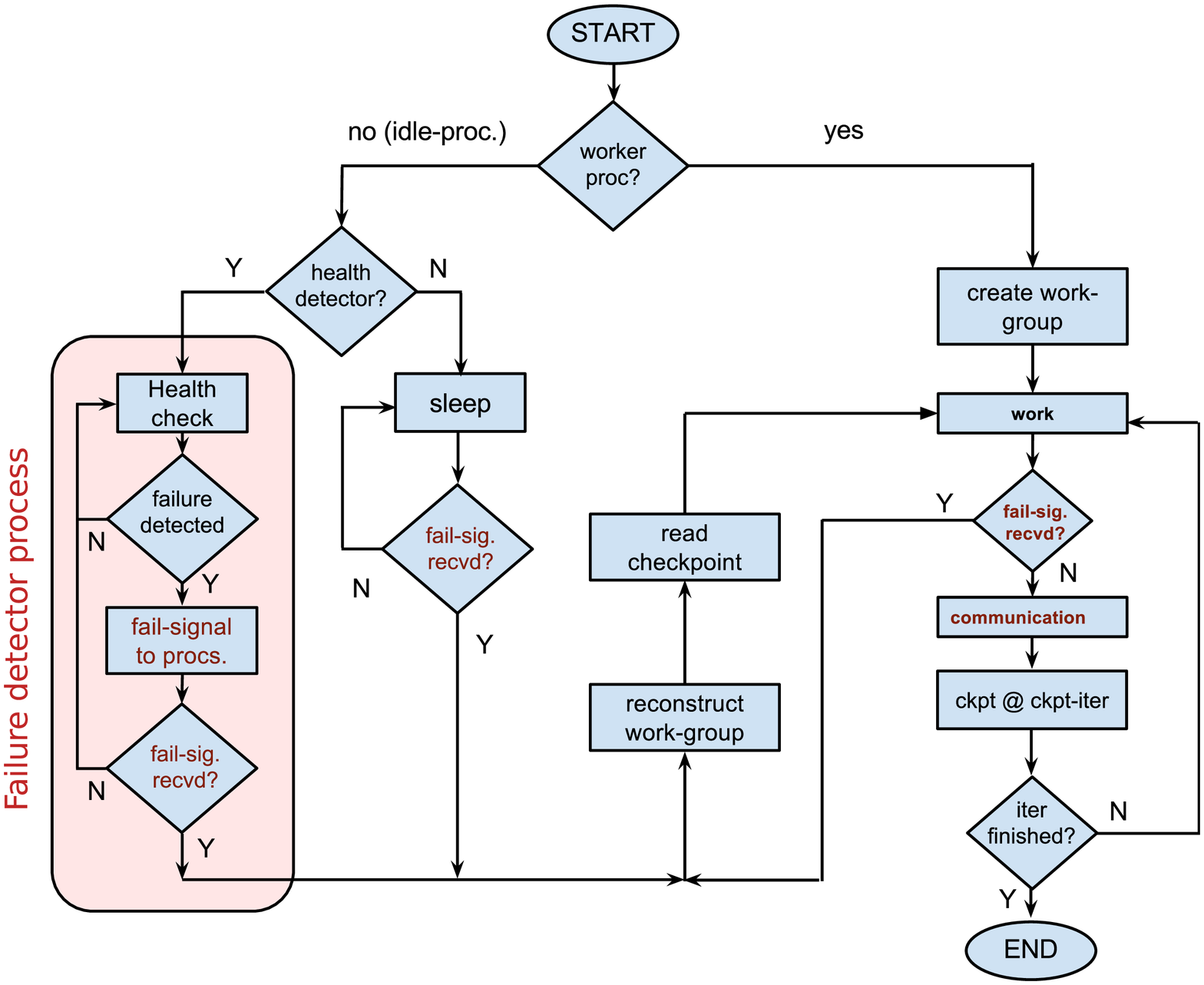}  
	\caption[]{Algorithm flow-chart with spare processes that are kicked-in in the main program in case of failures.
	\label{chart:flowchart-FT-program}}\vspace{.00cm}
\end{figure}

A flowchart of the modified algorithm is shown in
Fig. \ref{chart:flowchart-FT-program}. At the start of the
application, the processes are categorized into working and idle
processes. One of the idle processes acts as an FD process, whereas
the rest stay idle. The worker processes form a worker group
and perform computation. The communication routines are
checked for a failure acknowledgment signal from the FD process. The
FD process periodically checks the global state of the processes. Upon
detection of failure(s), it informs all remaining processes which then
enter the recovery stage of the algorithm to reconstruct the worker
group and restart from a consistent checkpoint. The FD process
itself joins the worker group if no idle process is further
available.

The current scheme has following restrictions which we plan to address 
in the future work.
\begin{enumerate} 
\item{The number of failures a program can sustain is equal to the
    number of idle processes specified at the start of the program.}
\item{The fault tolerance capability of a program ends if the FD process
    becomes a worker or encounters a failure itself.}
\item{Only those network failures can be 
detected that can be uniformly seen by the effected processes as well 
as by the FD process.}
\end{enumerate}

\subsection{Fault Tolerance Overhead}

Each of the fault tolerance techniques carries a certain amount of
overhead in terms of time and/or resources. Our approach requires the
allocation of some extra nodes for fallback scenarios, which is a
resource overhead. The calculation of the optimal number of extra
nodes for a particular case depends on several factors including job
size, job duration, the MTTF of the system, etc. and is out of scope for
this paper. In the following, we term overhead as the increase in
application run time due to its fault tolerance capabilities.

In principle, the fault detection with global ping messages is an
expensive operation. In our method, a FD process performs one-sided
pings to get the global health view. On the other hand, the worker
processes check for a failure acknowledgment signal from the FD
process before each communication call. Thus, from the working
processes' standpoint, the failure detection mechanism adds very
little cost to the overall run-time of the algorithm. As we shall see
in Sect. \ref{sec:results}, this cost is negligible.

The first major overhead is introduced by checkpointing. There
can be two kinds of checkpoints: a global PFS-level checkpoint, and a
neighbor level checkpoint. In case of a restart, the data is
initialized from a consistent checkpoint. The processes must redo the
work, up to the point where the actual failure happened. This
is the source of the second overhead and depends on the instance where the
failure occurred between two checkpoints.

In case of a failure, all processes go through the failure detection
and recovery stage of the program. This constitutes the third form of
overhead and can be decomposed into three categories:
\begin{itemize} 
\item{Failure detection overhead/communication with failed
    processes\,(OH$_{F1}$): This is the time it takes for the FD
    process to successfully detect and acknowledge the failures to
    other processes. Meanwhile, the healthy processes trying to
    communicate with the failed processes end up in 
    timeout-based returns unless a failure acknowledgment is received.}
\item{Rebuilding of work group\,(OH$_{F2}$): This step involves the
    creation of a new worker group, replacing failed processes
    (shown in Listing. \ref{lst:reconstruct_comm_worker}). Due to the
    blocking nature of the \textit{gaspi\_group\_commit()} procedure, this
    overhead is non-negligible.}
\item{Reinitialization of data\,(OH$_{F3}$): After the new work group
    is formed, the data structure gets initialized from the last
    consistent checkpoint. If the checkpoint is not available on the
    local node, it is fetched from the neighbor node of the failed
    process by the checkpoint library.}
\end{itemize}

In the next sections, we perform benchmarks and get a quantitative
notion of the above mentioned overheads in a practical scenario.

%% file: experimental_framework.tex
\section{Experimental Framework}
\label{sec:experimental_framework}

We use the Lanczos algorithm to demonstrate the usage of our fault
tolerance technique.  The Lanczos algorithm is an iterative scheme to
find eigenvalues of a sparse matrix. We use it to
find the low-lying eigenvalues of a test matrix.  Listing
\ref{pseudocode:lanczos-algo} shows the pseudo-code of the algorithm.
Each iteration calculates the new Lanczos vectors, $\alpha$,
and $\beta$. After obtaining the $\alpha$ and $\beta$ values of each
iteration, the approximated minimum eigenvalues are determined using
the QL method and are checked against a convergence criterion.  The
checkpointing data consists of two consecutive Lanczos vectors,
$\alpha$, and $\beta$.

\begin{algorithm}
\begin{algorithmic}
\For{j:=1,2, ..., ConvergenceCriterion}
                \Function {lanczos-step}{}
                        \State ${\omega}_j \gets A{\nu}_j$
                        \State ${\alpha}_j \gets {\omega}_j . {\nu}_j$
                        \State ${\omega}_j \gets {\omega}_j - {\alpha}_j{\nu}_j - {\beta}_j{\nu}_{j-1}$
                        \State ${\beta}_{j+1} \gets \|{\omega}_j\|$
                        \State ${\nu}_{j+1} \gets {\omega}_j / {\beta}_{j+1} $
                \EndFunction

                ${CalcMinimumEigenVal()}$
\EndFor
\end{algorithmic}
\caption{The Lanczos algorithm for finding eigenvalues of a matrix $A$}\label{pseudocode:lanczos-algo}
\end{algorithm}

For a parallel spMVM operation, the pre-processing stage includes the
setup of communication. In this stage, the spMVM operation is divided
into two parts, a local part for which the process has right-hand
vector values\,(RHS) locally available, and a remote part for which
the process needs to fetch RHS values from other processes. In the
pre-processing stage, each process determines the indices of the RHS
that it needs from other processes. These indices are communicated to
the respective processes, who then write (via one-sided GASPI
communication) the RHS values of those indices before every spMVM
iteration.

In the fault tolerant version of the Lanczos algorithm, each process
writes a checkpoint after the pre-processing stage. This checkpoint
stores information relevant for communication with other
processes. After failure recovery, the rescue process reads the
checkpoint of the failed process.  In this way, the rescue process is
informed about the communicating partners and the respective RHS
indices to communicate to other processes. Every non-failing process
also refreshes its list of communication partners by replacing the
rank of the failed process with the new rescue process.  Using this
method, the program can resume the computations after failure recovery
without having to perform the pre-processing step again (which would add
the overall cost of recovery).

In addition to the main Lanczos application, the underlying spMVM library 
also needs to be prepared for fault tolerance. Each blocking communication 
call in spMVM library now performs a check for the failure acknowledgment signal. 
After the processes detect a failure signal from the FD process, no further 
communications are performed.

The matrix for our benchmark case arises from the quantum-mechanical 
description of electron transport properties in graphene. Graphene is 
a blueprint for quasi two-dimensional materials with many interesting 
properties and prospective applications in several fields of nanotechnology 
and nanoelectronics. A matrix generation library tool is used to construct 
the matrix on the fly. Depending upon the specified geometry size, each 
process allocates its own chunk of the matrix. This way, the expensive 
step of reading the matrix from PFS is avoided.

\textit{Testbed:} All tests were done on the LiMa 
cluster\footnote{{LiMa} cluster at the {E}rlangen {R}egional {C}omputing 
{C}enter ({RRZE}): \url{http://www.hpc.rrze.fau.de/systeme/lima-cluster.shtml}} 
at RRZE, whose nodes are equipped with two Intel Xeon 5650 ``Westmere" chips 
with a base frequency of 2.66\,GHz.
Each node has 24\,GB of RAM\,(12\,GB per NUMA domain). The system has 
Mellanox QDR InfiniBand\,(IB) and GBit Ethernet interconnects.

%% file: results.tex
\section{Results}
\label{sec:results}
In this section, we test our benchmark application with various
runtime scenarios and compare their relative overhead. 
We have varified the recovery mechanism by killing the 
application processes in follwing three ways: 
i) Exiting the processes using `exit(-1)' 
within the program ii) Using `kill -9 $<$process-id$>$' 
iii) Physically introducing a network failure. 
Turning off complete compute nodes was not done 
due to limitations of the batch queuing system
which would delete the complete job owing to the dead node.

In the Lanczos algorithm, the stopping
criterion depends on finding a range of eigenvalues up to a certain
accuracy. For benchmarking purposes, we use a fixed number of
iterations ($3500$) as the stopping criterion. The neighbor-node level
checkpoints are taken at every $500$th iteration. As discussed in the
previous section, the matrix checkpoint is only stored once after its
pre-processing stage at the start of the algorithm. The periodic
checkpoints consist of only Lanczos vectors and the calculated
eigenvalues up to the corresponding iteration. The Graphene matrix
consists of $1.2 \cdot 10^8$ rows and columns, and $1.5 \cdot 10^9$ 
non-zeros. The global size of each checkpoint is
$\approx1.9$\,GBytes. The GASPI communication timeout value is set 
to 1 second, after which the processes retry communication unless 
a failure acknowledgement is recieved from the FD process. Furthermore, the FD 
process performs the ping scan on all processes every 3 seconds.

\begin{figure}[tb]
	\centering
	\includegraphics[width=8cm,clip=true]{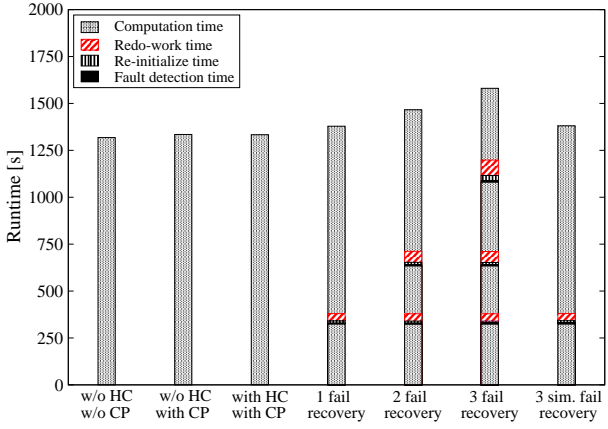}
	\caption[]{Various runtime scenarios of Lanczos application on 256 nodes. Each failure recovery cost $\approx17$ seconds.
	\label{chart:FT_1-2-3-FailRecovery_256-nodes}}\vspace{.00cm}
\end{figure}

Figure \ref{chart:FT_1-2-3-FailRecovery_256-nodes} shows the 
runtime of the Lanczos benchmark on 256 nodes 
(256 processes, 12 threads/process) in several cases. 
For this benchmark, the processes are killed using 
`exit(-1)' at a specific iteration in order to have a deterministic 
redo-work time. 
The application is started with four idle processes (nodes)
reserved for failure recovery. The first case represents the
application runtime where no health check is performed and no
checkpoints are taken (`w/o HC, w/o CP' bar). This runtime marks the
baseline case of our overhead study. The neighbor-node level
checkpoints add negligible overhead of around 0.01\% (`w/o HC,
with CP' bar). Furthermore the case of health check adds no additional
overhead (`with HC, with CP' bar). Significant overhead only appears in
case of failure recovery scenarios. Each failure adds approximately
$64$ seconds to the total runtime of the application. The average
failure detection takes $\approx7$ seconds on 256 nodes. The
`re-initialization' overhead, which includes group reconstruction and
checkpoint reading causes $\approx10$ seconds on average. Thus a
failure detection and recovery costs up to around $17$ seconds. The
redo-work time constitutes a major part of the total overhead. The
average time for redo-work is the time between two successive
checkpoints. Owing to a good checkpoint strategy with very low
overhead, the checkpoint frequency can be increased which will lead to
the reduction of redo-work time. Recovery from more failures adds
approximately proportional overhead as shown in
Fig. \ref{chart:FT_1-2-3-FailRecovery_256-nodes} with two and three
recovery case runtimes. In practice, a likely scenario is to have
multiple failures simultaneously (e.g., failure of a node with
multiple processes). Thus, we have used a threaded FD process (with 8
threads) to check the state of more than one process by monitoring
one-sided pings in parallel on different communication queues. This is
highlighted in `3 sim. fail recovery' case, where three simultaneous
failures are detected with the overhead of one failure detection.

It is important to observe and analyse the scaling behavior of a
fault detection mechanism. Table \ref{table:detection_time} shows 
the ping scan time of the FD process without any failure. 
The FD process takes approximately $1ms$ to perform a ping with 
each healthy process. The table also shows the failure detection and 
acknowledgement time in case of a failure. For this benchmark, 
one randomly selected process is terminated using 
`kill -9 $<$process-id$>$' at a random instance during 
the application run. The time is measured between the instance of 
fault injection and the acknowledgement of failures by the 
processes. The experiments show a good scalability of 
fault detection and acknowledgement method upto 256 nodes.  
It is important to mention that this time is the summation 
of the fault detecion overhead OH$_{F1}$ (as explained in previous section)
as well as the delay of 3 seconds between two ping scans.

\begin{table}
\begin{center}
        \begin{tabular}{|C{1.8cm} |C{.6cm}| C{.6cm} | C{.6cm} | C{.6cm} | C{.6cm} | C{.6cm} |}
        \hline
        Num. of Nodes	& 8 & 16 & 32 & 64 & 128 & 256 \\ \hhline{|=|=|=|=|=|=|=|} 
	Avg. ping scan time[s] & 0.010 & 0.018 & 0.036 & 0.067 & 0.129 & 0.255 \\ \hline   
        Failure detection and  ack. time[s] &4.9 $\pm$0.7 & 5.3 $\pm$0.7& 5.5 $\pm$0.8& 4.3 $\pm$0.7 & 5.7 $\pm$1.2 & 5.3 $\pm$0.8   \\ \hline
    \end{tabular}
\caption{The average ping scan time of the FD process and the failure detection time (and standard deviation using 10 runs) with respect to the number of nodes.}
\label{table:detection_time}
\end{center}
\end{table}

%% file: related_work.tex
\section{Related Work}
\label{sec:related_work}

The prototype implementation of ULFM using OpenMPI has produced 
similar work to ours in terms of failure recovery strategies and/or
performance evaluation.

Bland et al. have implemented a fault tolerant Monte Carlo
Communication Kernel proxy-app using ULFM in \cite{exampi14}. They
have demonstrated two recovery modes of the application, a
classic checkpoint/restart approach and an alternative solution 
in which the critical data is stored on neighboring 
nodes after each iteration. The
recovery is done by spawning an additional process which reads the
data from the failed process's neighboring node. In
\cite{fault-tolerant-montecarlo}, a different approach of data
recovery is implemented to develop a fault tolerant Multi-Level Monte
Carlo based application using ULFM. The data from failed processes are
altogether discarded and only survivors' data is used to complete the
application further. Depending on the number of failures, the
estimation error deteriorates the quality of results. A similar strategy 
was pursued by Ali et al. to develop a 2D PDE solver in
\cite{application-level-fault-recovery}. They compare three alternative
data recovery approaches based on C/R and approximation
techniques. All mentioned studies have found ULFM to have
acceptable overhead for failure recovery. In
\cite{evaluating-ULFM-for-mpi-applications}, Laguna et al. have
conducted a broad and critical evaluation of ULFM. They have first
discussed preferable recovery modes based on the nature of the
applications and then performed a case study on a ULFM-based
implementation of a fault-tolerant molecular dynamics application
(ddcMD). They have found the time of revoking and shrinking the
communicator to be increasing linearly with increasing number of
nodes. In the end, they have suggested improvements to ULFM to make it
an attractive option for application developers.

The ULFM studies are based on the detection of faults by communication 
failure between processes. No explicit failure detection
methods are used. In \cite{Assessing-FD-for-mpi-jobs}, K. Kharbas et
al. have evaluated two kinds of fault detector mechanisms for MPI
applications based on periodic and sporadic probing. The probing is
based on ping-pong style messaging. They have only reported the
overhead for failure-free cases which ranges from 1\% to 21\%, averaging
around 10\% for NAS-parallel benchmarks. The fault detector overhead
is negligible when separate background processes are used to check
failures on the same set of nodes and the main application. In this
case the background processes use Gigabit Ethernet instead of the
Infiniband network used by the application, but this is not a suitable
design as a large category of faults are network related. They
conclude that a separate periodic fault detection mechanism is a
superior method as compared to an sporadic approach.

%% file: summary.tex
\section{Summary}
\label{sec:summary}

We have implemented a fault-tolerant version of a parallel
Lanczos algorithm using the GPI-2 communication library 
with custom extensions. An explicit failure detector (FD) was 
designed to monitor the
health of all processes. The FD process performs periodic one-sided
pings to monitor the health state of all processes. We have
used the idea of spare processes in combination with a
checkpoint/restart mechanism for the replacement of failed
processes. One of these spare processes is designated to be the FD
process, whereas the rest stays idle until a failure occurs and they
join the worker processes. The efficiency of checkpoints is optimized
by using a neighbor node-level checkpoint/restart library which is
also fault tolerance aware.

The benchmark performed on 256 nodes show that having an explicit FD
causes no overhead to the working processes.  The recovery from a
single failure causes a total overhead of around 64 seconds, a large
part which comes from the redo-work time.  The failure detection
itself takes $\approx$7 seconds, whereas the communication
reconstruction and data re-initialization from the checkpoint add
$\approx$10 seconds of overhead. Each additional failure adds similar
cost. On the other hand, if multiple failures happen
simultaneously (likely scenario for node failures), they have the
potential to be detected at the cost of a single failure. 
Our failure detection mechanism shows good scaling behaviour on the 
scaling test performed up to 256 nodes. 

\textit{Discussion and Future work:}
The fault tolerance research is in its infancy stages regarding
application-driven fault tolerance. Only having FT communication is
not nearly enough. Many other components are 
needed to build an FT application in its true sense, e.g., 
failure information propagation to healthy processes, 
communication reconstruction after failure, data recovery method, etc. 
These add extra burden in terms of application development.

More fault tolerance related functionalies can be included within or
on top of the GASPI framework in order to simplify the development of
fault tolerant codes. These include failed processes acknowledgment by
groups, the replacement of failed processes by a newly spawned ones, etc. 
Furthermore, functionality can be provided to
assign the identity of the failed process to the newly spawned/replaced
process.
 
Our future work is targeted to remove the current limitations of 
our approach and making the technique more general and user
friendly. The redundancy approach can be implemented to make the
FD process fault tolerant. We also plan to compare this fault
tolerance approach with the Open MPI's ULFM functionality.